\documentclass[twocolumn,10pt,amsmath,amssymb]{revtex4}

\usepackage{amssymb}
\usepackage{epsf}
\usepackage{bm}

\newcommand{\n}[1]{n_\mathrm{#1}}
\newcommand{\Ge}{\Gamma_\mathrm{e}}
\newcommand{\G}[1]{\Gamma_\mathrm{#1}}
\newcommand{\Zmid}{\left<Z\right>}
\newcommand{\ZZmid}{\left<Z^2\right>}
\newcommand{\kB}{k_\mathrm{B}}
\newcommand{\aB}{a_{ij}^\mathrm{B}}

\begin{document}

\title{Nuclear fusion reaction rates for strongly coupled ionic mixtures}
\author{A.~I.~Chugunov}
\affiliation{Ioffe Physical-Technical Institute, Politekhnicheskaya
26, 194021 Saint-Petersburg, Russia}

\author{H.~E.~DeWitt}
\affiliation{Lawrence Livermore National Laboratory, Livermore, CA
94550, USA}

\begin{abstract}
We analyze the effect of plasma screening on nuclear reaction rates
in dense matter composed of atomic nuclei of one or two types. We
perform semiclassical calculations of the Coulomb barrier
penetrability taking into account a radial mean field potential of
plasma ions. The mean field potential is extracted from the results
of extensive Monte Carlo calculations of radial pair distribution
functions of ions in binary ionic mixtures. We calculate the
reaction rates in a wide range of plasma parameters and approximate
these rates by an analytical expression that is expected to be
applicable for multicomponent ions mixtures. Also, we analyze
Gamow-peak energies of reacting ions in various nuclear burning
regimes. For illustration, we study nuclear burning in
$^{12}$C--$^{16}$O mixtures.
\end{abstract}

\maketitle

\section{Introduction}
\label{s:introduct}

Nuclear fusion in dense stellar matter is most important for the
evolution of all stars. The main source of energy for main-sequence
stars is provided by hydrogen and helium burning. Subsequent burning
of carbon and heavier elements \cite{clayton83} drives an ordinary
star along the giant/red-giant branch to its final moments as a
normal star. Nuclear burning is also important for compact stars. It
is the explosive burning of carbon and other elements in the cores
of massive white dwarfs; it triggers type Ia supernova explosions
(see, e.g., \cite{hoeflich06} and references therein). Explosive
burning in the envelopes of neutron stars can produce type I X-ray
bursts \cite{sb06} and superbursts (that are observed from some
X-ray bursters; e.g., Refs.\ \cite{cummingetal05,guptaetal06}).
Accreted matter which penetrates into deeper layers of the neutron
star crust undergos pycnonuclear reactions. They can power thermal
radiation observed from neutron stars in soft X-ray transients in
quiescent states (see, e.g., Refs.\
\cite{YL03,Transient03,Transient04,YLG05,guptaetal06,pgw06,lh07,shternin07,YakPycno06}).
Thus, nuclear reactions are important in stars at all evolutionary
stages.

It is well known that nuclear reaction rates in dense matter are
determined by astrophysical $S$-factors, which characterize nuclear
interaction of fusing atomic nuclei, and by Coulomb barrier
penetration preceding the nuclear interaction. We will mostly focus
on the Coulomb barrier penetration problem. Fusion reactions in
ordinary stars proceed in the so called classical thermonuclear
regime in which ions (atomic nuclei) constitute a nearly ideal
Boltzmann gas. In this case the Coulomb barrier between reacting
nuclei is almost unaffected by plasma screening effects produced by
neighboring plasma particles. The Coulomb barrier penetrability is
then well defined.

However, in dense matter of white dwarf cores and neutron star
envelopes the ions form a strongly non-ideal Coulomb plasma, where
the plasma screening effects can be very strong. The plasma
screening greatly influences the barrier penetrability and the
reaction rates. Depending on density and temperature of the matter,
nuclear burning can proceed in four other regimes \cite{svh69}. They
are the thermonuclear regime with strong plasma screening, the
intermediate thermo-pycnonuclear regime, the thermally enhanced
pycnonuclear regime, and the pycnonuclear zero-temperature regime.
The reaction regimes will be briefly discussed in Sec.\
\ref{reactreg}. In these four regimes the calculation of the Coulomb
barrier penetration is complicated. There have been many attempts to
solve this problem using several techniques but the exact solution
is still a subject of debates (see
\cite{Gasques_etal05,YGABW06,OCP_react} and reference therein).

In our previous paper \cite{OCP_react} we studied nuclear reactions
in a one component plasma (OCP) of ions. Now we extend our
consideration to binary ionic mixtures(BIMs). In Sec.\
\ref{PhysCond} we discuss physical conditions and reaction regimes.
In Sec.\ \ref{meanfield} we describe the results of our most
extensive and accurate Monte Carlo calculations, analyze them and
use to parameterize the mean-field potential. In Sec.\ \ref{enh}
this potential is employed to calculate the enhancement factors of
nuclear reaction rates. We study the Gamow peak energies and the
effects of plasma screening on astrophysical $S$-factors in Sec.\
\ref{sec:Gamow}. Section \ref{sec:res} is devoted to the analysis of
the results. We conclude in Sec.\ \ref{sec:conc}. In the Appendix we
suggest a simple generalization of our results to the cases of weak
and moderate Coulomb coupling of ions.

\section{Physical conditions and reaction regimes}
\label{PhysCond}

\subsection{Nuclear reaction regimes}
\label{reactreg}

We consider nuclear reactions in dense matter of white dwarf cores
and outer envelopes of neutron stars. This matter contains atomic
nuclei (ions, fully ionized by electron pressure) and strongly
degenerate electrons, which form an almost uniform background of
negative charge. For simplicity, we will not consider the inner
crust of neutron stars (with density higher than the neutron drip
density $\sim 4\times 10^{11}$~g\,cm$^{-3}$ \cite{NV73}), which
contains also free degenerate neutrons.

We will generally study a multicomponent mixture of ions $j=1, 2,
\ldots$ with atomic mass numbers $A_j$ and charge numbers $Z_j$. The
total ion number density can be calculated as $\n{i}=\sum_j\, n_j$,
where $n_j$ is a number density of ions $j$.  It is useful to
introduce the fractional number $x_j=n_j/\n{i}$ of ions $j$. Let us
also define the average charge number
 $\langle Z \rangle=\sum_j\, x_j Z_j$
and mass number
 $\langle A \rangle=\sum_j\, x_j A_j$
of ions. The charge neutrality implies that the electron number
density is $n_e=\langle Z \rangle \n{i}$. The electron contribution
to the mass density $\rho$ can be neglected, so that
$\rho\approx\langle A \rangle m_\mathrm{u} n $, where
$m_\mathrm{u}\approx 1.66054\times 10^{-24}$~g is the atomic mass
unit.

The importance of the Coulomb interaction for ions $j$ can be
described by the coupling parameter $\Gamma_j$
\begin{eqnarray}
   \Gamma_j=\Gamma_{jj}=\frac{Z_j^2\,e^2}{a_j \kB T}, \quad
   a_j=Z_j^{1/3}a_\mathrm{e},\quad a_\mathrm{e}=\left(\frac{3}{4\pi
   n_\mathrm{e}}\right)^{1/3},
\end{eqnarray}
where $a_j$ and $a_\mathrm{e}$ are the ion-sphere and
electron-sphere radii, $\kB$ is the Boltzmann constant, and $T$ is
the temperature. Note, that the electron charge within an ion sphere
exactly compensates the ion charge $Z_je$; $\Gamma_j$ gives the
ratio of characteristic electrostatic energy $Z_j^2e^2/a_j$ to the
thermal energy $\kB T$.

If $\Gamma_j\ll 1$, the ions constitute an almost ideal Boltzman
gas, while for $\Gamma_j\gtrsim 1$ they are strongly coupled by
Coulomb forces and constitute either a Coulomb liquid or solid. The
transformation from the gas to the liquid at $\Gamma_j\sim 1$ is
smooth, without any phase transition. According to highly accurate
Monte Carlo calculations, a classical OCP of ions solidifies at
$\Gamma\approx 175$ (see, e.g.\ \cite{DWSBY2001}).

We will discuss fusion reactions
\begin{equation}
 (A_i,\ Z_i)+(A_j,\ Z_j) \rightarrow
 (A^\mathrm{comp}_{ij},\ Z^\mathrm{comp}_{ij}),
\end{equation}
where $A^\mathrm{comp}_{ij}=A_i+A_j$ and
$Z^\mathrm{comp}_{ij}=Z_i+Z_j$ refer to a compound nucleus. The
reaction rate is determined by an astrophysical $S$-factor. For a
non-resonant reaction, it is a slowly varying function of energy
(see, e.g., Ref.\ \cite{YGABW06}). It is determined by the
short-range nuclear interaction of fusing nuclei and by the Coulomb
barrier penetration problem. The latter task can be reduced to the
calculation of the contact probability $g_{ij}(0)$, which is the
value of the quantum-mechanical pair correlation function of
reacting nuclei $i$ and $j$ at small interionic distances
$r\rightarrow 0$. Finally, the reaction rate can be written as (see
Sec.\ \ref{sec:Gamow} and \cite{ichimaru93})
\begin{equation}
    R_{ij}=\frac{2\,\aB }{\left(1+\delta_{ij}\right)\pi\hbar}\,\,n_i\,n_j
    S_{ij}\left({E^\mathrm{pk}_{ij}}^\prime\right) g_{ij}(0),
    \label{R_throw_g}
\end{equation}
where
\begin{equation}
    S_{ij}(E)= \sigma_{ij}(E) E
                \exp(2\pi\eta_{ij})
\end{equation}
is the astrophysical factor. It should be taken at an appropriate
energy $E={E^\mathrm{pk}_{ij}}^\prime$, which is the ``Gamow peak
energy'', corrected for the plasma screening (see Sec.\
\ref{sec:Gamow} and Fig.\ \ref{fig:gamow}); $\delta_{ij}$ is
Kronecker delta, that excludes double counting of the same
collisions in reactions of identical nuclei ($i=j$). For these
reactions, $\aB=\hbar^2/\left(2\mu_{ij}\,Z_i\,Z_j\,e^2\right)$
reduces to the ion Bohr radius. Furthermore,
$\mu_{ij}=m_\mathrm{u}\,A_iA_j/A^\mathrm{comp}_{ij}$ is the reduced
mass of the nuclei and $\sigma_{ij}(E)$ is the fusion cross section.
Finally, $\eta_{ij}=Z_iZ_j e^2\, \sqrt{\mu_{ij}/(2E\hbar^2)}$
determines  the penetrability of the Coulomb barrier.

\begin{figure}[t!]
    \begin{center}
        \leavevmode
        \epsfxsize=80mm \epsfbox{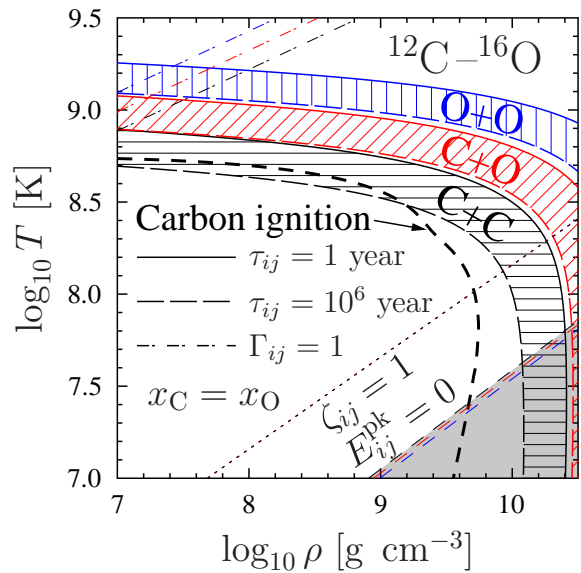}
    \end{center}
    \caption{(Color online)    Temperature-density diagram
    for $^{12}$C+$^{16}$O mixture
    with equal number densities of $^{12}$C and $^{16}$O.
    The dash-dot lines correspond (from top to bottom) to
    $\Gamma_\mathrm{OO}$, $\Gamma_\mathrm{CO}$,
    $\Gamma_\mathrm{CC}=1$. For temperatures below these lines,
    the effects of plasma screening on corresponding reaction rates become strong.
    The dotted line is
    $\zeta_{ij}=1$. This line is the same
    for all $i$ and $j$ because of the same $A/Z=2$ ratio.
    The reacting nuclei become
    bound along the dashed straight lines
    (the Gamow-peak energy passes through zero, see Sec.\
    \ref{sec:Gamow}).
    For lower temperatures (gray region) the mean field model (used in this paper)
    becomes inadequate.
    The shaded regions are bound by solid and long-dashed lines
    of constant burning times
    ($1$~yr and $10^{6}$ yr). The thick dash line is the carbon
    ignition curve (Sec.\ \ref{sec:res}).
      }
    \label{Fig_ReacRegimes}
\end{figure}

Figure \ref{Fig_ReacRegimes} is the temperature-density diagram for
a $^{12}$C+$^{16}$O mixture with equal number densities of both
species. Also, we present the lines of constant dimensionless
parameters (see Sec.\ \ref{dimlesparams}) $\Gamma_{ij}=1$ (dash-dot
lines) and $\zeta_{ij}=1$ (the dotted line). The dash-dot lines
(from top to bottom) correspond to $\Gamma_\mathrm{OO}=1$,
$\Gamma_\mathrm{CO}=1$, and $\Gamma_\mathrm{CC}=1$. The
$\zeta_{ij}=1$ line is the same for all combinations of ions because
of equal $A/Z=2$ rations for $^{12}$C and $^{16}$O ions. The shaded
regions are bounded by the solid and long-dashed lines of constant
burning times of $1$ and $10^6$~yr, respectively. The region for the
O+O reaction is highest, for C+O intermediate, and for C+C lowest
(as regulated by the height of the Coulomb barrier). The lines of
constant burning time are similar. At $\rho \lesssim
10^{9}$~g\,cm$^{-3}$ they mainly depend on temperature because of
exponentially strong temperature dependence of the reaction rates.
These parts of the lines correspond to thermonuclear burning (with
weak or strong plasma screening). With growing density, the curves
bend and become nearly vertical (that corresponds to the
pycnonuclear burning due to zero-point vibrations of atomic nuclei;
see \cite{svh69,vhs67,sk90,YGABW06}). The pycnonuclear reaction rate
is a rapidly varying function of density, which is either slightly
dependent or almost independent of temperature.

The carbon ignition curve (discussed in Sec.\ \ref{sec:res}) is
shown by the thick dashed line.

\subsection{Dimensionless parameters}
\label{dimlesparams}

Let us consider a binary mixture of ions with mass numbers $A_1$ and
$A_2$, charge numbers $Z_1$ and $Z_2$, and fractional numbers
$x_1=n_1/n_\mathrm{i}$ and $x_2=1-x_1$. Following Itoh et al.\
\cite{ikm90}, in addition to the ion radii $a_i=a_{ii}$ we introduce
an effective radius
\begin{equation}
\label{a_12_def}
   a_{12}=\frac{a_1+a_2}{2}
   =\frac{Z_1^{1/3}+Z_2^{1/3}}{2}\,a_\mathrm{e}
\end{equation}

It is also convenient to define the
parameter
\begin{equation}
    \Ge=e^2/(a_\mathrm{e}\,\kB\,T),
\end{equation}
The coupling parameters for ions can be written as
\begin{equation}
    \Gamma_i=\Gamma_{ii}=\G{e}\,Z_i^{5/3}.
\end{equation}
In addition, we introduce an effective Coulomb coupling parameter
\begin{equation}
    \G{12}=Z_1Z_2e^2/(a_{12}\kB\,T),
\end{equation}
to be used later for studying a reaction of the type
$(A_1,Z_1)+(A_2,Z_2)$.

Furthermore, we introduce the corresponding ion radius and the
Coulomb coupling parameter for a compound nucleus
$(A_i+A_j,Z_i+Z_j)=(A_{ij}^\mathrm{comp}, Z_{ij}^\mathrm{comp})$,
\begin{equation}
  a^\mathrm{comp}_{ij}=a_\mathrm{e}\,\left(Z^\mathrm{comp}_{ij}\right)^{1/3},\quad
  \Gamma^\mathrm{comp}_{ij}=\G{e}\left(Z^\mathrm{comp}_{ij}\right)^{5/3}.
\end{equation}

The Coulomb barrier penetrability in the classical thermonuclear
regime (neglecting plasma screening) is characterized by the
parameter
\begin{equation}
     \tau_{ij}=\left(\frac{27\pi^2 \mu_{ij} Z_i^2 Z_j^2 e^4}
                          {2\kB T\hbar^2}
               \right)^{1/3}.
               \label{tau_ij}
\end{equation}
The Gamow peak energy in this regime is
\begin{equation}
    E_{ij}^\mathrm{pk}=\kB T\,\tau_{ij}/3.
    \label{GamowEpk}
\end{equation}
The effects of plasma screening on the Gamow peak energy and
tunneling range are discussed in Sec.\ \ref{sec:Gamow}.

Also, we introduce a dimensionless parameter
\begin{equation}
   \zeta_{ij}= 3\,\Gamma_{ij}/\tau_{ij}.
\end{equation}
In an OCP, $\zeta_{ii}$ reduces to the parameter $\zeta$ defined in
Ref.\ \cite{OCP_react} and has a simple meaning -- it is the ratio
of the tunneling length of nuclei with the Gamow peak energy
(calculated neglecting of the plasma screening effects) to the ion
sphere radius $a_{ij}$.

\section{Mean field potential}
\label{meanfield}

Direct calculation of the contact probability $g_{ij}(0)$ is very
complicated even in OCP. In principle, it can be done by Path
Integral Monte Carlo (PIMC) simulations, but it requires very large
computer resources. PIMC calculations performed so far are limited
to a small number of plasma ions (current OCP results are described
in \cite{ogata97,pm04,mp05}). In BIMs one needs a larger number of
ions, $N_1$ and $N_2$, to get good accuracy.

We will use a simple mean-field model, summarized in this section.
It can be treated as a first approximation to the PIMC approach
\cite{aj78}.  Previously we showed \cite{OCP_react} it is in good
agreement with PIMC calculations of Militzer and Pollock \cite{mp05}
for OCP.

A classical  pair correlation function, calculated by the classical
Monte Carlo (MC) method, can be presented as
\begin{equation}
  g^\mathrm{MC}_{ij}(r)=
  \exp\left[\frac{1}{\kB T}\,
            \left(\frac{Z_iZ_j e^2}{r}-H_{ij}(r)\right)
     \right],\label{Hij}
\end{equation}
where $H_{ij}(r)$ is the mean field potential to be determined.

The Taylor expansion of $H_{ij}(r)$ in terms of $r$ for a strongly
coupled ion system should contain only even powers \cite{widom63}.
The first two terms can be derived analytically in the same
technique as proposed by Jancovici \cite{jancovici77} for OCP. The
mean field potential at low $r\ll a_{ij}$ takes form \cite{oii91}
\begin{equation}
    H_{ij}(r)\approx\,\kB T h_{ij}^0
    -\frac{Z_i Z_j
    e^2
    }{2a^\mathrm{comp}_{ij}}
      \left(\frac{r}{a^\mathrm{comp}_{ij}}\right)^2,
      \label{Low_r_Hij}
\end{equation}
where
\begin{equation}\label{h}
    h_{ij}^0
    =f_0(\Gamma_i)+f_0(\Gamma_j)-f_0(\Gamma_{ij}^\mathrm{comp}),
\end{equation}
and $f_0(\Gamma)$ is the Coulomb free energy per ion in OCP. Here,
the linear mixing rule is assumed. In principle, $h_{ij}^0$ can be
extrapolated from MC $H_{ij}(r)$ data (e.g., Ref.\ \cite{ds99}), but
the problem is delicate \cite{rosenfeld96} and we expect that using
linear mixing rule is preferable. The linear mixing rule is
confirmed with high accuracy \cite{dsc96,ds03} but its uncertainties
were a subject of debates;  see Ref.\ \cite{Potekhin_Mixt} for
recent results.

\begin{figure}[t!]
    \begin{center}
        \leavevmode
        \epsfxsize=80mm \epsfbox{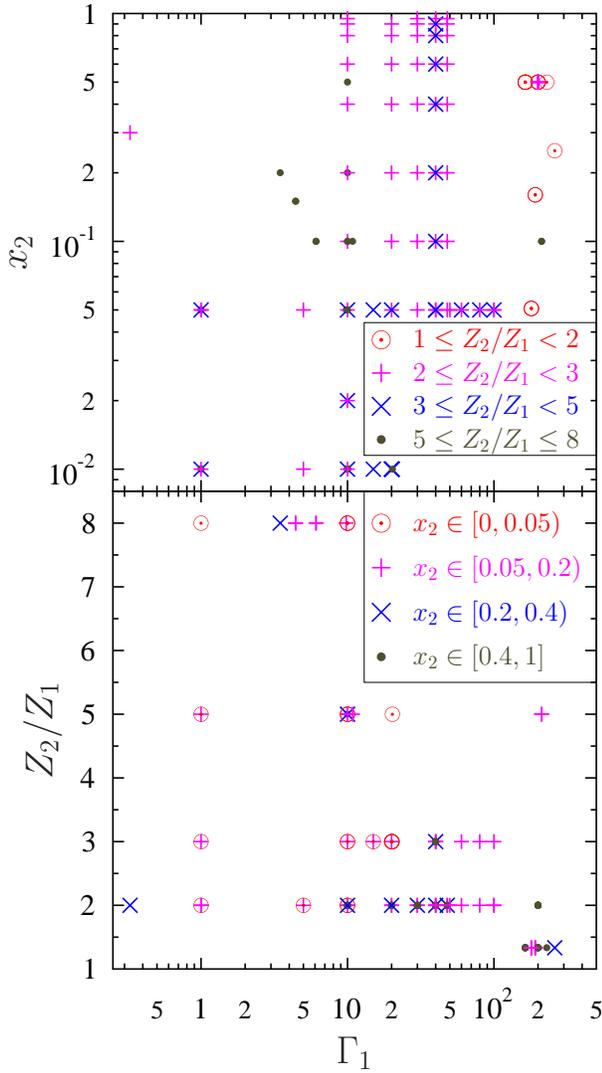}
    \end{center}
    \caption{(Color online) Every BIM MC run is shown on both panels.
    {\it Up:}
    $x_2$ versus $\G{1}$.
    Circles,  pluses, crosses and dots refer to
    runs with
    different ratios of $Z_2/Z_1$: $1\le Z_2/Z_1<2$,
    $2\le Z_2/Z_1<3$, $3\le Z_2/Z_1< 5$, $5\le Z_2/Z_1\le 8$,
    respectively.
    {\it Down:} $Z_2/Z_1$ versus $\G{1}$.
    Circles,  pluses, crosses and dots
    refer to runs with different fractions $x_2$:
    $x_2<0.05$, $0.05\le x_2<0.2$, $0.2\le x_2< 0.4$, $0.4\le Z_2/Z_1\le 1$.
      }
    \label{Fig_Our_data}
\end{figure}
\begin{figure}[h!]
    \begin{center}
        \leavevmode
        \epsfysize=193mm \epsfbox{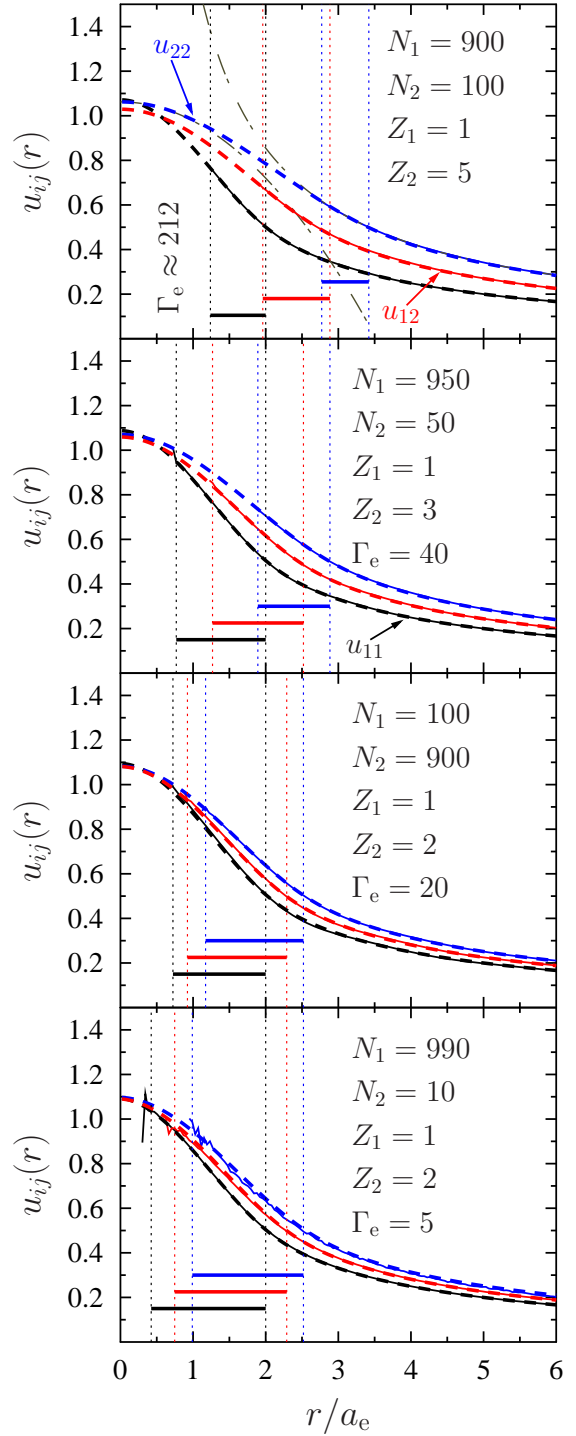}
    \end{center}
    \caption{(Color online) Mean field potential $u_{ij}$ versus $r/a_\mathrm{e}$
    for four MC runs. Solid lines are extracted from MC data,
    dashed lines are given by the fit (\ref{u_fit}). On each plot
    the lines, which are upper at  $r\gtrsim 2\,a_\mathrm{e}$,
    correspond to $u_{22}(r)$; lower lines are $u_{12}$
    and lowest lines give $u_{11}$.
    Thin vertical dotted lines connected by thick solid lines show the ranges of $r$,
    which are used in the fitting.}
    \label{Fig:U_r}
\end{figure}

To determine the mean field potential in BIMs we performed extensive
MC calculations of $H_{ij}(r)$. We did 129 MC runs. In each run we
calculated pair correlation functions $g_{11}$, $g_{12}$ and
$g_{22}$ for a given set of the parameters $Z_1$, $Z_2$, $N_1$,
$N_2$, and $\Gamma_1$.  In Fig.\ \ref{Fig_Our_data} each set is
shown by a symbol. The top panel demonstrates a fraction of
particles 2, $x_2$, versus $\Gamma_1$. Circles, pluses, crosses and
dots refer to different ratios of $Z_2/Z_1$: $1\le Z_2/Z_1<2$, $2\le
Z_2/Z_1<3$, $3\le Z_2/Z_1< 5$, $5\le Z_2/Z_1\le 8$, respectively. In
the bottom panel we show the $Z_2/Z_1$-ratio versus $\Gamma_1$.
Circles,  pluses, crosses and dots correspond to different values of
$x_2$: $x_2<0.05$, $0.05\le x_2<0.2$, $0.2\le x_2< 0.4$, $0.4\le
x_2\le 1$. For some parameter sets (e.g., for $\Gamma_1=180$,
$x_2\approx0.05$, $Z_2/Z_1=4/3$) we use a number of MC runs with
different initial configurations (random or regular lattice) and/or
different simulation times. We did not find any significant
dependence of the mean field potential (for $r<2a_{ij}$) on the
initial parameters. From all our simulations we extract $H_{11}(r)$,
$H_{12}(r)$ and $H_{22}(r)$ and remove some low-$r$ points with
large MC noise. We then fit these functions at $r\lesssim 2a_{ij}$
by a simple universal analytical expression
\begin{eqnarray}\label{u_fit}
    H_{ij}(r)
    &=&\kB T\, \Gamma_{ij}^\mathrm{eff} \\
    &\times&\sqrt{
      \frac{U_0^2+U_1\,x^4+U_2\,x^8}
           {1+U_3\,x+U_4\,x^2+U_5\,x^4+U_6\,x^6+U_2\,x^{10} }
         },
\nonumber
\end{eqnarray}
where $x=r/a^\mathrm{eff}_{ij}$,
$U_0=h_{ij}^0/\Gamma^\mathrm{eff}_{ij}$, $U_1=0.0319$, $U_2=0.0024$,
$U_3=0.22\left(\Gamma^\mathrm{eff}_{ij}\right)^{-0.7}$,
$U_4=2^{-1/3}
\left(a^\mathrm{eff}_{ij}/a^\mathrm{comp}_{ij}\right)^2/U_0$,
$U_5=0.111$, and $U_6=0.0305$. Note, that $U_0$ and $U_4$ are fixed
by the low-$r$ asymptote of $H_{ij}(r)$ at strong Coulomb coupling,
$\Gamma_{ij}\gg1$, Eq.\ (\ref{Low_r_Hij}). For a better
determination of $H_{12}(r)$, we introduce the effective ion sphere
radius $a^\mathrm{eff}_{ij}$ and the Coulomb coupling parameter
$\Gamma^\mathrm{eff}_{ij}$:
\begin{equation}
   a^\mathrm{eff}_{ij}=\left(\frac{a_i^p+a_j^p}{2}\right)^{1/p},\quad
    \Gamma^\mathrm{eff}_{ij}=\frac{Z_iZ_je^2}{\kB
    T a^\mathrm{eff}_{ij}}.
\label{a_eff}
\end{equation}
Here, $p$ is an additional fit parameter which is found to be
$p=1.6$. Note, that $a^\mathrm{eff}_{ii}=a_{ii}=a_i$ for reactions
with identical nuclei ($i=j$). If, however, $i\ne j$ the parameters
$a^\mathrm{eff}_{ij}$ and $a_{ij}$ differ by a few percent and this
difference is very important. Using $a_{ij}$ to define $x$ in Eq.\
(\ref{u_fit}) and using $H_{ij}(r)$ to reproduce the pair
correlation function $g_{ij}(r)$ through Eq.\ (\ref{Hij}), we find
the first peak of $g_{ij}$ to be unrealistically deformed for large
$Z_2/Z_1\gtrsim5$.

Let us mention that in the OCP case our fit expression (\ref{u_fit})
is not the same as we used previously \cite{OCP_react}. Equation
(\ref{u_fit}) seems to be better because it better describes the
first peak of $g(r)=g_{ii}(r)$, especially at very large
$\Gamma\gtrsim 200$. For an OCP at  $\Gamma\lesssim 200$, the old
and new fit expressions give approximately the same accuracy of
$H(r)=H_{ii}(r)$.

Four examples of MC runs are shown in Fig.\ \ref{Fig:U_r}. For each
run (on each plot) we show dimensionless mean field potential
$u_{ij}=H_{ij}(r)/(\kB T\Gamma_{ij})$ versus $r/a_\mathrm{e}$. Solid
lines are MC data, dashed lines are derived from the fit expression
(\ref{u_fit}). One can see numerical noise of MC data at low $r$; it
is especially strong in the lowest plot for $\Gamma_\mathrm{e}=5$.
During fitting procedure we removed some low-$r$ points with high MC
noise. Our fitting ranges are shown by pairs of thin dotted lines
connected by a thick solid line. The low-$r$ bound of the fitting
region is different for $H_{11}$, $H_{12}$, and $H_{22}$ and for
each MC run. The high-$r$ bound was taken to be $2a_{ij}$ for all
data. Thus, all our fits include the vicinity of the first peak of a
pair correlation function at $r\sim 1.8 a_{ij}$. Although we have no
MC data at low $r$, we expect that our approximation is well
established because it satisfies the accurate low-$r$ asymptote
(\ref{Low_r_Hij}). Also, it satisfies the large-$r$ asymptote
$u_{ij}(r\rightarrow 1)=a_{ij}/r$, which corresponds to the fully
screened Coulomb potential. Both asymptotes for $u_{22}$ are shown
on the upper plot by dash-dot lines. One can see, that they nicely
constrain the mean field potential.

A poor MC statistics at low $r$ can be, in principle, increased
using specific MC schemes \cite{cg03} but this is beyond the scope
of the present paper.

Previously the mean field approximations was used by Itoh et al.\
\cite{ikm90} and Ogata et al.\ \cite{oiv93}. While we use the
correct form of the linear mixing rule (\ref{h}) to calculate
$u_{12}(0)$, the cited authors restricted themselves by OCP results,
which depend only on $\Gamma_{12}$ (but not on $Z_2/Z_1$). The
results of this simplification are discussed in Sec.\ \ref{enh};
they are visible in Fig.\ \ref{Fig:H_r}. In addition, our
approximation is based on a more representative set of MC runs (see
Fig.\ \ref{Fig_Our_data}), and each run was done with better
accuracy. For example, where is no visible noise $g_{ij}(r)$-noise
in the vicinity of the first peak in out data, whereas this noise is
obvious in Fig.\ 1 of Ref.\ \cite{oiv93}. We have also checked that
our approximation (\ref{u_fit}) better reproduces all our MC data
than approximations of both groups \cite{ikm90,oiv93}. As noted by
Ogata et al.\ \cite{oiv93}, the height of the first $g_{ij}(r)$-peak
depends on the fraction of ions with larger charge (the peaks become
higher with increasing this fraction). Our data confirm this
statement for a not very large Coulomb coupling parameter,
$\Gamma_1\lesssim 10$. Ogata et al.\ \cite{oiv93} added a correction
for this feature in their approximation of $u_{11}$. It helps to
describe the first $g_{11}(r)$-peak for some of our runs (e.g., for
$\Gamma_1=10$, $Z_2/Z_1=5$ and $x_2=0.2$). However, we think that we
still do not have enough data to quantitatively describe this
correction to the linear mixing with good accuracy, and we use the
mean field potential (\ref{u_fit}) based on the linear mixing rule
throughout the paper. In the Appendix we describe the corrections to
the linear mixing on a phenomenological level. 

\section{Enhancement of nuclear reaction rates}\label{enh}

\begin{figure*}
    \begin{center}
        \leavevmode
        \epsfxsize=0.95\textwidth \epsfbox{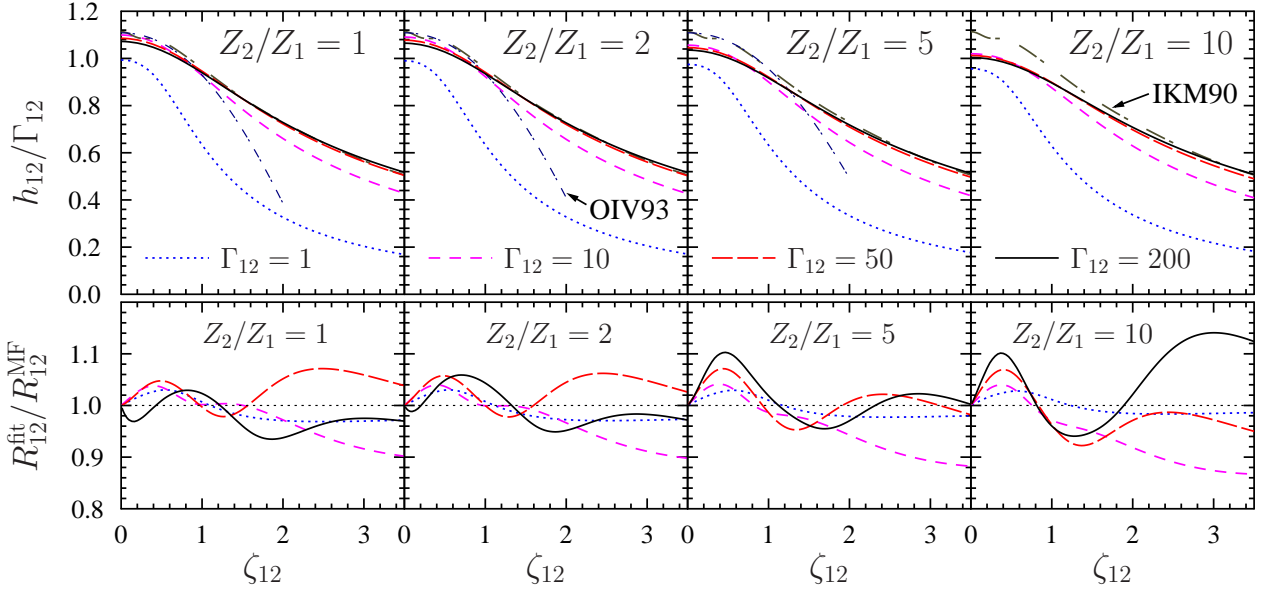}
    \end{center}
    \caption{(Color online) {\it Top:} Enhancement factors
    $h_{ij}/\Gamma_{12}$ for several
    $\Gamma_{12}$ (=1, 10, 50, 200; each shown by a line on every panel)
    and $Z_2/Z_1$ (=1, 2, 5, 10; each on its own panel).
    The long dash-dot line presents the fit expression by Itoh et
    al. \cite{ikm90} (IKM90); the short dash-dot lines are the results by
    Ogata et al. \cite{oiv93} (OIV93), calculated for
    $\Gamma_{12}=50$, as an example (the latter lines are shown only
    for $\zeta_{12}<2$ and $Z_2/Z_1\le5$ which are the fit bounds in OIV93).
    {\it Bottom:} Ratios of $R_{12}^\mathrm{fit}$, given by Eq.\ (\ref{h_fit}),
    to the mean field reaction rate $R_{12}^\mathrm{MF}$, given by
    (\ref{thermon}),
    versus $\zeta_{12}$. See text for details.}
    \label{Fig:H_r}
\end{figure*}

Let us introduce the enhancement factor $F^\mathrm{scr}_{ij}$ of a
nuclear reaction rate $R^\mathrm{scr}_{ij}$ under the effect of
plasma screening,
\begin{equation}
  R^\mathrm{scr}_{ij}=R^\mathrm{th}_{ij}\,F^\mathrm{scr}_{ij},\quad
  F^\mathrm{scr}_{ij}=\exp (h_{ij}).
  \label{h_def}
\end{equation}
Here, $R^\mathrm{th}_{ij}$ is the reaction rate in the absence of
the screening. In can be calculated using the classical theory of
thermonuclear burning,
\begin{equation}
    R^\mathrm{th}_{ij}=\frac{4 n_i
    n_j}{1+\delta_{ij}}\,\frac{S_{ij}(E^\mathrm{pk}_{ij})}{k_\mathrm{B}T}\,
    \sqrt{\frac{2E^\mathrm{pk}_{ij}}{3\mu_{ij}}}\exp\left(-\tau_{ij}\right).
\end{equation}
The barrier penetrability parameter $\tau_{ij}$ and the Gamow peak
energy $E^\mathrm{pk}_{ij}$, which enter this equation, are defined,
respectively, by Eqs.\ (\ref{tau_ij}) and (\ref{GamowEpk}).

Using $H_{ij}(r)$ we can calculate the reaction rate. This
calculation takes into account the plasma screening enhancement in
the mean-field approximation (as was done for OCP in our previous
paper \cite{OCP_react}),
\begin{eqnarray}
  R^\mathrm{MF}_{ij}&=&\frac{n_i\,n_j S_{ij}\left({E^\mathrm{pk}_{ij}}^\prime\right)}{1+\delta_{ij}}\,  
  \sqrt{8  \over \pi\mu_{ij} (\kB T)^3}\,
\nonumber \\
  && \times \int_{E_\mathrm{min}}^\infty {\rm d} E\,
  \exp\left[-\frac{E}{\kB T} -P_{ij}(E) \right].
\label{thermon}
\end{eqnarray}
Here, $E$ is the center-of-mass energy of the colliding nuclei (with
a minimum value $E_\mathrm{min}$ at the bottom of the potential
well), $\exp(-E/\kB T)$ comes from the Maxwellian energy
distribution of the nuclei, and
\begin{equation}
     P_{ij}(E)={ 2 \sqrt{2\mu_{ij}} \over \hbar} \,
     \int_{0}^{r_t} {\rm d}r \,
     \sqrt{{Z_iZ_j e^2 \over r} -H_{ij}(r) -E}
\label{P(E)}
\end{equation}
is the Coulomb barrier penetrability ($r_t$ being a classical
turning point).

In the WKB mean-field approximation the enhancement factor of the
nuclear reaction rate is
\begin{equation}
   F^\mathrm{scr}_{ij}=R^\mathrm{MF}_{ij}\{H_{ij}\}/R^\mathrm{MF}_{ij}\{H_{ij}=0\}.
\label{WKBenh}
\end{equation}
In the cases of physical interest, $R^\mathrm{MF}_{ij}\{0\}$ can be
integrated by the saddle-point method. However, we always integrate
in Eq.~(\ref{thermon}) numerically.

The calculated enhancement factor $F^\mathrm{scr}_{ij}$ can be
fitted as
\begin{equation}\label{h_fit}
    \log F^\mathrm{scr}_{ij}=f_0\left(\frac{\Gamma_i}{\tau_{ij}}\right)
            +f_0\left(\frac{\Gamma_j}{\tau_{ij}}\right)
            -f_0\left(\frac{\Gamma_{ij}^\mathrm{comp}}{\tau_{ij}}\right),
\end{equation}
where $f_0(\Gamma)$ is a free energy per ion in an OCP. We use the
most accurate available analytic fit for $f_0(\Gamma)$ suggested by
Potekhin and Chabrier \cite{pc00}:
\begin{eqnarray}
   f_0(\Gamma)&=&A_1\left[
                       \sqrt{\Gamma\,\left(A_2+\Gamma\right)}
               \right.       \nonumber \\
   & -& \left . A_2\ln\left(\sqrt{\Gamma/A_2}+\sqrt{1+\Gamma/A_2}\right)
                 \right]
   \nonumber \\
   &+&2 A_3 \left[\sqrt{\Gamma}-\arctan\left(\sqrt\Gamma\right)\right]
      \nonumber \\
   &+&B_1\left[\Gamma-B_2\ln\left(1+\frac{\Gamma}{B_2}\right)\right]
   \nonumber \\  & +&\frac{B_3}{2}\ln\left(1+\frac{\Gamma^2}{B_4}\right).
\label{krokodilische}
\end{eqnarray}
Here, $A_1=-0.907$, $A_2=0.62954$,
$A_3=-\sqrt{3}/2-A_1/\sqrt{A_2}\approx0.2771$, $B_1=0.00456$,
$B_2=211.6$, $B_3=-10^{-4}$, $B_4=0.00462$. We also introduce a
number of additional parameters:
\begin{eqnarray}
y_{ij}&=&\frac{4\,Z_i\,Z_j}{\left(Z_i+Z_j\right)^2},\quad
c_1=0.013\,y_{ij}^2,\nonumber \\
 c_2&=&0.406\,y_{ij}^{0.14}, \quad
c_3=0.062\,y_{ij}^{0.19}+1.8/\Gamma_{ij},
\nonumber \\
    t_{ij}&=&\left(1+c_1\,\zeta_{ij}+c_2\,\zeta_{ij}^2+c_3\zeta_{ij}^3\right)^{1/3}.
    \label{h_fit_addpar}
\end{eqnarray}

The root-mean-squared relative error of the fit of $F_\mathrm{scr}$
is $5\%$. The maximum error $\sim 14\%$ occurs at $\zeta_{ij}\approx
4.8$, $\Gamma_{12}\approx25$, and $Z_2/Z_1\approx10$. The fit was
constructed for $1\le Z_2/Z_1\le10$, $1\le\Gamma_{12}\le200$ (we
also use $\Gamma_{12}=400$ and $600$ for testing) and $\zeta_{12}\le
8$ (larger values of $\zeta_{12}$ were not included into fitting but
used for tests). As for OCP in our previous paper \cite{OCP_react},
at larger $\Gamma$ the relative errors increase.

In the limit of $\zeta_{ij}\rightarrow 0$ we have  $\log
F^\mathrm{scr}_{ij}=h_{ij}^0$. This is the well known relation
(e.g., \cite{salpeter54,dgc73}); its formal proof is given in
\cite{ys89}.

Our fit formula (\ref{h_fit}) has the same form as the thermodynamic
relation (\ref{h}), but contains Coulomb coupling parameters
$\Gamma_{ij}$ divided by $\tau_{ij}$. It generalizes our OCP result
\cite{OCP_react}.

Figure \ref{Fig:H_r} shows the normalized enhancement factor
$h_{12}/\Gamma_{12}$ (bottom panels) as a function of $\zeta_{12}$
for several values of $\Gamma_{12}$ ($=1$, $10$, $50$, $200$;
separate lines on each plot) and $Z_2/Z_1$ ($=1$, $2$, $5$ and $10$,
separate plots). The first panel (for $Z_2=Z_1$) corresponds not
only to a reaction in OCP, but to reactions of identical nuclei in a
mixture; the mean field potential and the plasma screening
enhancement do not depend on an admixture of other elements [see
Eqs.\ (\ref{u_fit}) and (\ref{h_fit})], as long as the Coulomb
coupling parameter is not too small, $\Gamma_{ij}\gtrsim 1$, and the
linear mixing is valid. The larger $\Gamma_{ij}$, the weaker is the
dependence of $h_{12}/\Gamma_{12}$ on $\Gamma_{ij}$. Long-dashed
lines, plotted for $\Gamma_{12}=50$, significantly differ from
dashed lines which are for $\Gamma_{12}=10$, but can be hardly
distinguished from solid lines, plotted for $\Gamma_{12}=200$.
Therefore, the total enhancement factor $\exp\left(h_{12}\right)$
depends exponentially on $\Gamma_{12}$ in the first approximation.
The normalized enhancement factor $h_{12}/\Gamma_{12}$ suggested by
Itoh et al.\ \cite{ikm90} (IKM90) is shown by long dash-dot lines.
It is independent of $\Gamma_{12}$, and we show one line in each
plot. The short dash-dot lines demonstrate the fit expression of
Ogata et al.\ \cite{oiv93} (OIV93) (for $\Gamma_{12}=50$ as an
example). The lines are cut at $\zeta_{12}=2$ (that bounds the fit
validity). We show no OIV93 line on the panel for $Z_2/Z_1=10$
because there are no OIV93 simulations for such large $Z_2/Z_1$
ratios. The two lines, IKM90 and OIV93, demonstrate higher
enhancement, especially for $\zeta_{12}\lesssim1$ and large
$Z_2/Z_1$. First, IKM90 and OIV93 used less accurate thermodynamic
approximations for calculating $u_{ij}(0)$. For large
$Z_2/Z_1$-ratios, their fit error increases because of using the OCP
expression (dependent only of $\Gamma_{12}$, but not on $Z_2/Z_1$)
for determining $u_{ij}(0)$. This increases a difference between our
and their approximations at large $Z_2/Z_1$ (see also a discussion
in Sec.\ IIIC of Ref.\ \cite{YGABW06}). IKM90 and our
$\zeta_{12}$-dependence of the enhancement factors is qualitatively
the same, but the results of OIV93 (and of the preceding paper
\cite{oii91}) are qualitatively different. As we show previously
\cite{OCP_react}, such results contradict recent PIMC results by
Militzer and Pollock \cite{mp05} for OCP. The deviation of the IKM90
enhancement factors at low $\zeta_{ij}$ values comes possibly from
using an oversimplified mean field potential.

The indicated differences between IKM90, OIV93 and our results can
lead to very large differences of the total enhancement factor
$\exp\left(h_{12}\right)$. For example, at $Z_2/Z_1=5$ and
$\Gamma_{12}=200$ the difference can reach five orders of magnitude.
Accordingly, we do not present the IKM90 and OIV93 results in lower
panels, which give the ratio of the reaction rates
$R^\mathrm{fit}_{12}$, given by the fit expression (\ref{h_fit}), to
the reaction rate $R^\mathrm{MF}_{12}$, calculated in the mean field
approximation using Eq.\ (\ref{thermon}). A very small difference
($\lesssim 10\%$), shown on these plots, is true only by adopting
the mean field potential (\ref{u_fit}). The uncertainties of the
reaction rates, which come from the uncertainties of the mean field
potential and inaccuracy of the mean field approximation, can be
larger.

Let us stress that Eq.\ (\ref{h_fit}) is derived for the case of
strong screening; it is invalid at $\Gamma_{12}\lesssim 1$, where
the adopted linear mixing rule fails. A simple
correction of the enhancement factor for weaker screening is
suggested in the Appendix.
It allows us to extend the results for the case of weaker Coulomb
coupling.

\section{Astrophysical S-factors: Gamow peak energies at strong screening}
\label{sec:Gamow}

In the previous section we focused on the plasma screening of the
Coulomb barrier penetration in dense plasma environment. Now we
discuss the effects of plasma screening on the astrophysical
$S$-factor, that describes nuclear interaction of the reacting
nuclei after the Coulomb barrier penetration. The latter effects are
not expected to be very strong because of not too strong energy
dependence of $S_{ij}(E)$ (e.g., Ref.\ \cite{Gasetal07}), but we
consider them for completeness.

The astrophysical $S$-factor describes the effects of short-range
nuclear forces and it should depend on the parameters of the
reacting nuclei just before a reaction event.  The energy
${E^\mathrm{pk}_{ij}}^\prime$ substituted into $S(E)$ in Eq.\
(\ref{thermon}) should be corrected for the mean field potential
created by other ions, ${E^\mathrm{pk}_{ij}}^\prime=E+H_{ij}(0)$.
This correction is obvious and has been used in calculations (e.g.,
\cite{itw03}), but its formal proof has not been published, to the
best of our knowledge. The reaction rate in the presence of plasma
screening can be calculated as
\begin{eqnarray}
    R_{ij}&=&\frac{n_in_j}{1+\delta_{ij}}\left(\frac{8}{\pi\mu_{ij}T^3}\right)^{1/2}
    \nonumber \\
    &\times&
    \int \widetilde\sigma_{ij}(E)E \exp\left(-\frac{E}{T}\right) \mathrm{d}E,
    \label{R_sigma}
\end{eqnarray}
where $\widetilde\sigma_{ij}(E)$ is the reaction cross section including
the screening effects. To calculate $R_{ij}$, let us remind the
barrier penetration model a with parameter-free model of nuclear
interaction gives a good description of reaction cross sections
\cite{Gasques_etal05}. In the absence of plasma screening at not too
high energies $E$ (where only the s-wave channel is important), this
model reduces to the WKB calculation of the penetrability through a
potential which is the sum of the Coulomb potential and the
potential $V^\mathrm{N}_{ij}(r,v_{ij}^2)$, that describes the
short-range nuclear interaction. The latter potential depends on the
local relative velocity $v(r.E)$ of the reactants given by
\begin{equation}
v^2=\frac{2}{\mu_{ij}}\left[E+H_{ij}(r)-\frac{Z_iZ_je^2}{r}-V^\mathrm{N}_{ij}(r,v^2)\right].
\end{equation}
Let us generalize this model by adding the screening potential
$H_{ij}(r)$ and write the cross section as
\begin{eqnarray}
 \widetilde \sigma_{ij}(E)&=&\frac{\pi}{k_{ij}^2}
  \exp\left[-\frac{2\sqrt{2\mu_{ij}}}{\hbar}\int_{r_\mathrm{tn}}^{r_\mathrm{t}}
        \mathrm{d} r
  \right. \nonumber \\
   &\times&\left.   \sqrt{\frac{Z_iZ_je^2}{r}-H_{ij}(r)
                          +  V^\mathrm{N}_{ij}(r,v_{ij}^2)-E
                     }\right].
                     \label{sigma_tild}
\end{eqnarray}
Here, $r_\mathrm{tn}$ and $r_\mathrm{t}$ are classical turning
points and  $k_{ij}^2=2\mu_{ij} E/\hbar^2$. Substituting this
equation into Eq.\ (\ref{R_sigma}) we have
\begin{eqnarray}
    R&=&\frac{n_in_j}{1+\delta_{ij}}\left(\frac{8}{\pi\mu_{ij}T^3}\right)^{1/2}
    \nonumber \\
    &\times&
    \int \widetilde S_{ij}(E)\exp\left[-\frac{E}{T}-P_{ij}(E)\right] \mathrm{d}E,
\end{eqnarray}
where $\widetilde S_{ij}(E)$ is the astrophysical factor, calculated
in the presence of the plasma screening,
\begin{equation}
    \widetilde S_{ij}(E)=\widetilde \sigma_{ij} E
    \exp\left[P_{ij}(E)\right];
\label{S_tild}
\end{equation}
$P_{ij}(E)$ is defined by Eq.\ (\ref{P(E)}). The potential
$V^\mathrm{N}_{ij}(r,v)$ is non-zero only for very small $r\lesssim
r_\mathrm{N}$. Let us substitute Eqs.\ (\ref{P(E)}) and
(\ref{sigma_tild}) into (\ref{S_tild})  and divide all integrals
into two parts, at $r<r_\mathrm{N}$ and $r\geq r_\mathrm{N}$. The
integrals over $r>r_\mathrm{N}$ which come from $P_{ij}(E)$ and
$\widetilde\sigma_{ij}(E)$ will be exactly the same
($V^\mathrm{N}_{ij}(r,v)=0$ for  $r>r_\mathrm{N}$) and cancel each
other. As a result,
\begin{eqnarray}
    \widetilde S_{ij}(E)&=&\frac{\pi\hbar^2}{2\mu_{ij}}
    \exp\left\{\frac{2\sqrt{2\mu_{ij}}}{\hbar}\right.
    \nonumber \\
    &\times& \left[\int_0^{r_\mathrm{N}}
    \sqrt{\frac{Z_iZ_je^2}{r}
                            -H_{ij}(r)-E
                        }\, \mathrm d r\right.
    \label{S_tild2}  \\
    &-&\left. \left. \int_{r_\mathrm{tn}}^{r_\mathrm{N}}
        \sqrt{\frac{Z_iZ_je^2}{r}-H_{ij}(r)
                            +  V^\mathrm{N}_{ij}(r,v^2)-E
                     }\, \mathrm{d} r
                     \right]\right\}.
\nonumber
\end{eqnarray}
As expected, the astrophysical $S$-factor is determined by the
behavior of the total potential
$Z_iZ_je^2/r-H_{ij}(r)+V^\mathrm{N}_{ij}(r,v^2)$ in the low-$r$
region $r<r_\mathrm{N}$, where the nuclear forces are important.
Because $r_\mathrm{N}$ is much smaller than the ion sphere radius
$a_{ij}$, which is a typical length scale  of $H_{ij}(r)$ (e.g.,
Fig.\ \ref{Fig:U_r}), we can neglect variations of $H_{ij}(r)$ for
$r<r_\mathrm{N}$ and replace $H_{ij}(r)$ under the integrals in Eq.\
(\ref{S_tild2}) by $H_{ij}(0)$. The $S_{ij}(E)$-factor in the
absence of plasma screening is given by the same Eq.\
(\ref{S_tild2}) but at $H_{ij}(r)=0$. Thus, $\widetilde S_{ij}(E)$
is defined by the same equation as $S_{ij}(E^\prime)$, provided
$E^\prime=E+H_{ij}(0)$. In other words, $\widetilde
S_{ij}(E)=S_{ij}(E+H_{ij}(0))$ and the reaction rate is
\begin{eqnarray}
    R&=&\frac{n_in_j}{1+\delta_{ij}}\left(\frac{8}{\pi\mu_{ij} \kB^3 T^3}\right)^{1/2}
    \nonumber \\
    &\times& \int S_{ij}(E+H_{ij}(0))\exp\left[-\frac{E}{\kB T}-P_{ij}(E)\right]
    \mathrm{d}E.
    \label{R_fin}
\end{eqnarray}

If this simple barrier penetration theory were invalid and the
hypothesis of low-energy hindrance of nuclear reactions
\cite{jiangetal07} were correct, such a simple correction of
astrophysical $S$-factors for the plasma screening effects could be
insufficient, and the calculation of the reaction rate would be much
more complicated. We will not consider this possibility.

As in of the absence of the plasma screening, the main contribution
to the integral (\ref{R_fin}) comes from a narrow energy range. We
neglect the energy dependence of $S_{ij}(E+H_{ij}(0))$ in this range
and take the $S$-factor out of the integral. The contact probability
in the mean field model is given by
\begin{equation}
  g_{ij}(0)=\frac{2\sqrt{2\pi\mu_{ij}}}{\kB^{3/2} T^{3/2}}\frac{Z_1\,Z_2e^2}{\hbar}
  \int \exp\left[-\frac{E}{\kB T}-P_{ij}(E)\right] \mathrm{d}E,
\end{equation}
so that the reaction rate reads
\begin{equation}
    R_{ij}=\frac{2\,n_in_j}{1+\delta_{ij}}
    \frac{a^\mathrm{B}_{ij}}{\hbar\pi} S_{ij}
    \left(E^\mathrm{pk}_{ij}+H_{ij}(0)\right)
    g_{ij}(0).
\label{R_g_0}
\end{equation}
\begin{figure}
    \begin{center}
        \epsfxsize=0.45\textwidth \epsfbox{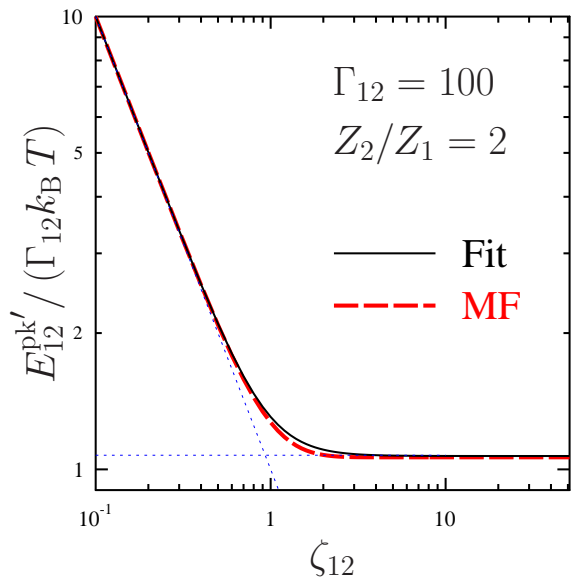}
    \end{center}
    \caption{(Color online) The Gamow peak energy
    ${E^\mathrm{pk}_{ij}}^\prime$ normalized for $\Gamma_{12}$
    as function of $\zeta_{12}$ at $Z_2/Z_1=2$ and $\Gamma_{12}=100$. Solid line
    is the fit expression (\ref{Epk}), dashed line shows the mean field result.
    Thin dotted lines correspond to the low-$\zeta_{ij}$ thermonuclear asymptote
    ${E^\mathrm{pk}_{ij}}^\prime=
    \Gamma_{12}/\zeta_{12}$ and high-$\zeta_{ij}$ pycnonuclear asymptote
    ${E^\mathrm{pk}_{ij}}^\prime=H_{ij}(0)$. See text for details.}
    \label{fig:Epk}
\end{figure}

\begin{figure*}[t!]
    \begin{center}
        \leavevmode
        \epsfxsize=0.95\textwidth \epsfbox{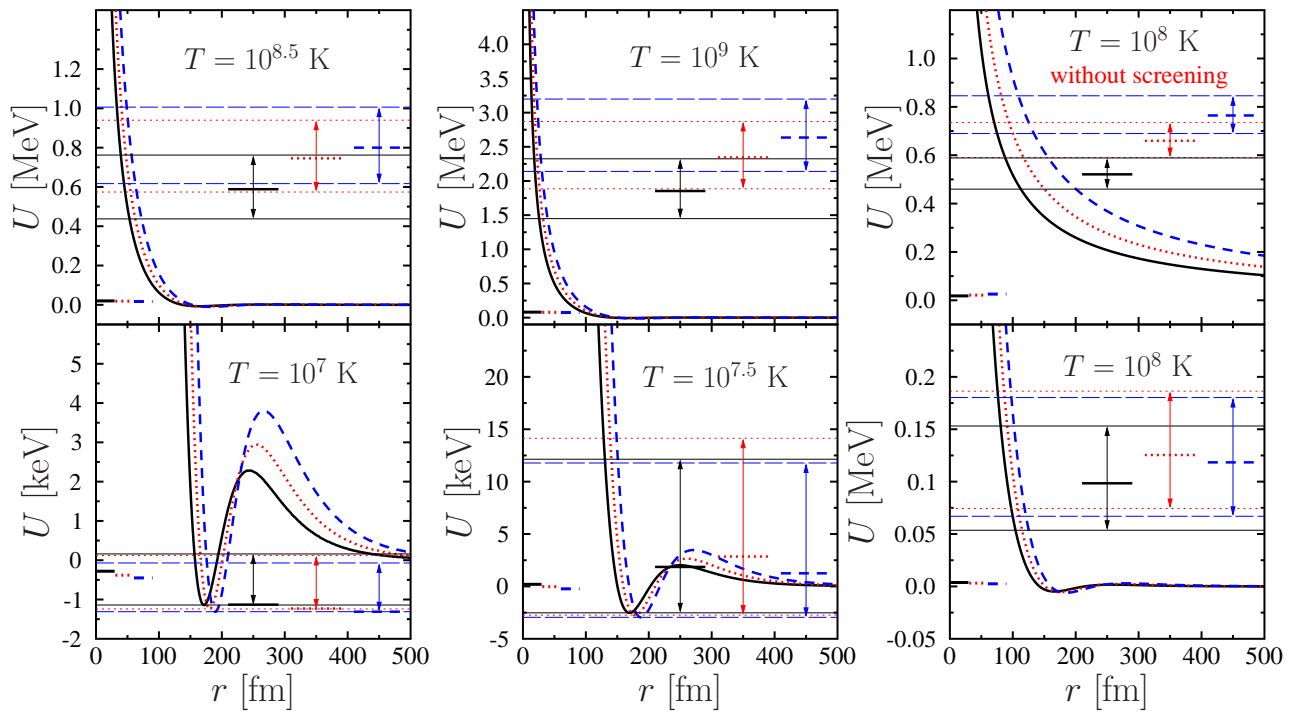}
    \end{center}
    \caption{(Color online) Effective mean-field Coulomb potentials $U_{ij}(r)$
    for the reactions $^{12}\mathrm{C}+^{12}\mathrm{C}$ (solid line),
    $^{12}\mathrm{C}+^{16}\mathrm{O}$ (dotted line) and
    $^{16}\mathrm{O}+^{16}\mathrm{O}$ (long-dash line) in the
    mixture of $^{12}\mathrm{C}$ and$^{16}\mathrm{O}$ with equal
    number densities of these nuclei at $\rho=5\times 10^9$~g~cm$^{-3}$
    and  five temperatures ($\log_{10} T[K]=9$, $8.5$, $8$, $7.5$, and $7$).}
    \label{fig:gamow}
\end{figure*}

We have calculated the modified Gamow-peak energy
${E^\mathrm{pk}_{ij}}^\prime=E^\mathrm{pk}_{ij}+H_{ij}(0)$, which
should be used as an argument of $S$-factors, in a wide range of
plasma parameters. It can be fitted as
\begin{equation}
   {E^\mathrm{pk}_{ij}}^\prime
   =\left[H_{ij}^3(0)+\left(\kB T\,\frac{\Gamma_{ij}}{\zeta_{ij}}\right)^3\right]^{1/3}.
\label{Epk}
\end{equation}
The maximum relative fit error for $1\le\Gamma_{ij}\le 200$,
$\zeta_{ij}\le 8$ and $0.1\le Z_i/Z_j\le 10$ is 4\%. It takes place
at $\Gamma_{ij}=1$, $\zeta_{ij}=1$ and $Z_i=Z_j$. Because of large
nuclear physics uncertainties in our knowledge of $S(E)$ at low
energies of astrophysical interest, this accuracy is more than
sufficient. An example of the dependence of
${E^\mathrm{pk}_{12}}^\prime/\Gamma_{12}$ on $\zeta_{12}$ is shown
in Fig.\ \ref{fig:Epk} for $Z_2/Z_1=2$ and $\Gamma_{12}=100$. The
solid line is the fit (\ref{Epk}), the dashed line is a result of
the exact ${E^\mathrm{pk}_{ij}}^\prime$ calculation in the mean
field model. The peak energy ${E^\mathrm{pk}_{ij}}^\prime$ has two
asymptotes shown by thin dotted lines. At low $\zeta\ll 1$, the
reaction occurs in the thermonuclear regime,
${E^\mathrm{pk}_{ij}}^\prime=\Gamma_{ij}/\zeta_{ij}$, which is the
standard classical result. For large $\zeta_{ij}\gg 1$, we have
${E^\mathrm{pk}_{ij}}^\prime \to H_{ij}(0)$.
The classical asymptote
${E^\mathrm{pk}_{ij}}^\prime=\Gamma_{ij}/\zeta_{ij}$ is applicable
at $\zeta_{ij}\lesssim 0.5$, where the plasma screening enhancement
can be very strong (tenths orders of magnitude). This is because in
the thermonuclear regime the tunneling length is not very large and
$H_{ij}(r)$ does not significantly change it (see Fig.\
\ref{fig:gamow}); ions tunnel through the Coulomb potential, shifted
by $H_{ij}(0)$. The shift increases the probability of close ions
collisions by a factor of $\exp\left(H_{ij}(0)/T\right)$, but does
not change the Maxwellian energy distribution and the dependence of
the tunneling probability on tunneling length. As a result, the main
contribution to the reaction rate comes from ions with the same
tunneling length (and the same ${E^\mathrm{pk}_{ij}}^\prime$), as in
the absence of plasma screening.

The large-$\zeta_{ij}$ asymptote is simple. For such parameters,
thermal effects are small and the ions, which mainly contribute to
the reaction rate, correspond to the minimum energy of the total
potential $Z_1Z_2/r+H_{ij}(r)$. This energy is small compared with
$H_{ij}(0)$, so that ${E^\mathrm{pk}_{ij}}^\prime \approx
H_{ij}(0)$. One can see a small difference of $H_{ij}(0)$, shown by
the thin dotted line in Fig.\ \ref{fig:Epk}, and the asymptote of
$H_{ij}(0)$ at large $\zeta_{ij}$. This difference does not
introduce significant uncertainties to the reaction rate because of
much larger nuclear-physics uncertainties caused by our poor
knowledge of $S(E)$ at low energies (e.g., Ref.\ \cite{YGABW06}). We
assume that the ${E^\mathrm{pk}_{ij}}^\prime$ approximation
(\ref{Epk}) is valid (at least qualitatively) not only for
thermonuclear burning with strong screening, but also for
pycnonuclear burning. In the pycnonyclear regime, the ions occupy
their ground states and oscillate near their lattice sites. The
reaction rates are determined by zero-point vibrations of the ions.
The energy of zero-point vibrations is typically small compared to
the minimum energy of the potential $Z_i Z_j/r+\widetilde
H_{ij}(\bm{r})$ (the latter is much lower than $\widetilde
H_{ij}(0)$, where $\widetilde H_{ij}(\bm{r})$ is an anisotropic
effective potential created by neighboring ions). Therefore, the
ions start with a small kinetic energy (the minimum of the $Z_i
Z_j/r+\widetilde H_{ij}(\bm{r})$) and tunnel to $r\rightarrow 0$
through the Coulomb potential plus the $\widetilde H_{ij}(\bm{r})$.
During tunneling, they fall into the potential well $\widetilde
H_{ij}({\bm r})$, and their energy increases by $\widetilde
H_{ij}(0)$. Therefore, ${E^\mathrm{pk}_{ij}}^\prime$ should be equal
to $\widetilde H_{ij}(0)$, but not to energy of zero-point
vibrations, as assumed in \cite{svh69,sk90,YGABW06}. Note, that in
the relaxed-lattice approximation $\widetilde H_{ij}(0)=H_{ij}(0)$
\cite{svh69,sk90}.

For example, let us consider $E^\mathrm{pk}_{ij}$ and the
characteristic (half-maximum) energy widths of the Gamow peak for a
$^{12}$C and $^{16}$O mixture (with equal number densities of C and
O ions) at $\rho=5 \times 10^9$ g~cm$^{-3}$. In Fig.\
\ref{fig:gamow} we show the effective total radial mean-field
Coulomb potentials $U_{ij}(r)=Z_iZ_je^2/r-H_{ij}(r)$ ($i,j=^{12}$C
or $^{16}$O) for five temperatures, $T=10^9$, $10^{8.5}$, $10^8$,
$10^{7.5}$, and $10^7$~K. For this mixture, the ion-sphere radius of
$^{12}$C ions is $a_\mathrm{C}$=98~fm, and it is
$a_\mathrm{O}$=108~fm for $^{16}$O ions. The solid, dotted and
long-dash lines show the CC, CO and OO potentials, respectively.
Each potential $U_{ij}(r)$ has a minimum at $r_{ij} \approx 2a_{ij}$
due to the Coulomb coupling. The thin horizontal lines connected by
double-arrow lines in Fig.\ \ref{fig:gamow} show the Gamow-peak
energy ranges. The types of thin lines are the same as for
$U_{ij}(r)$. Thick sections of short horizontal lines, which
intersect double-arrow lines, demonstrate the Gamow-peak energies
$E^\mathrm{pk}_{ij}$. Thick parts of short-dashed lines at small
$r<100$~fm indicate the thermal energy level $\kB T$ measured from
the bottom of $U_{ij}(r)$.

The top right panel marked ``without screening''\ shows the same
lines (as other panels) for $T=10^8$~K neglecting plasma screening.
Dramatic difference of unscreened and screened potentials and
corresponding Gamow-peak regions is obvious from comparison with
bottom panels. The plasma screening reduces the Gamow-peak energy
$E^\mathrm{pk}_{ij}$ by a factor of four, but (as described above)
does not affect significantly the Gamow-peak width.

Let us discuss Fig.\ \ref{fig:gamow} in more detail. The panels
plotted for $T=10^{8.5}$ and $10^{9}$~K (top left and top center,
respectively) refer to the thermonuclear reaction regime with strong
plasma screening.  The next two panels (right bottom and center
bottom) are for a colder plasma ($T=10^{8}\mbox{ K}$ and
$T=10^{7.5}\mbox{ K}$, respectively), while the last (left bottom)
panel is for a very cold plasma ($T=10^{7}\mbox{ K}$) (that is
certainly in the zero-temperature pycnonuclear regime). When the
temperature decreases, the Gamow-peak energy range becomes thinner
(note the difference of energy scales in different panels) and
shrinks to lower energies. If $T\gtrsim 10^8$~K, the Gamow peak
range is still at $E>0$ [belonging to continuum states in a
potential $U_{ij}(r)$], and the peak energy $E^\mathrm{pk}_{ij}$
(the short-dash horizontal line) is in the center of the Gamow-peak
range. The energies within this range are much higher than $\kB T$
supporting the statement that the main contribution to reaction
rates at sufficiently high $T$ comes from suprathermal ions. In
these cases, the underlying mean-field WKB approximation is expected
to be adequate. In the forth panel, the lowest energies of the
Gamow-peak range become negative (drop to bound states) and
$E^\mathrm{pk}_{ij}$ moves from the center to the lower bound of the
Gamow-peak range (so that the Gamow peak becomes significantly
asymmetric). The mean-field WKB approach based on the spherically
symmetric mean field potential $H_{ij}(r)$ may be still
qualitatively correct but becomes quantitatively inaccurate. Note
that the $U_{ij}(r)$-potential becomes positive for $r\gtrsim
200$~fm. This feature is determined by the right wing of the first
$g_{ij}(r)$-peak, where $g_{ij}(r)$ becomes smaller than one. This
wing is unimportant in the mean field model; our approximation gives
qualitatively correct $U_{ij}(r)$ for $r\gtrsim 2a_{ij}$
($g_{ij}(r)\rightarrow 1$ for $r\rightarrow \infty$). For the lowest
temperature in Fig.\ \ref{fig:gamow} the Gamow-peak energy range
fully shrinks to bound-state energies and the formal Gamow-peak
energy becomes lower than $\kB T$, nearly reaching the lower bound
of the Gamow-peak region. The mean-field WKB approximation breaks
down at these low temperatures, and the formally calculated
$E^\mathrm{pk}_{ij}$ is inaccurate. Nevertheless, the energy in the
argument of the $S$-factor,
${E^\mathrm{pk}_{ij}}^\prime=E^\mathrm{pk}_{ij}+H_{ij}(0)\approx
H_{ij}(0)$, is expected to be well defined.

\section{Results and discussion}
\label{sec:res}


\begin{figure}
    \begin{center}
        \epsfxsize=0.47\textwidth \epsfbox{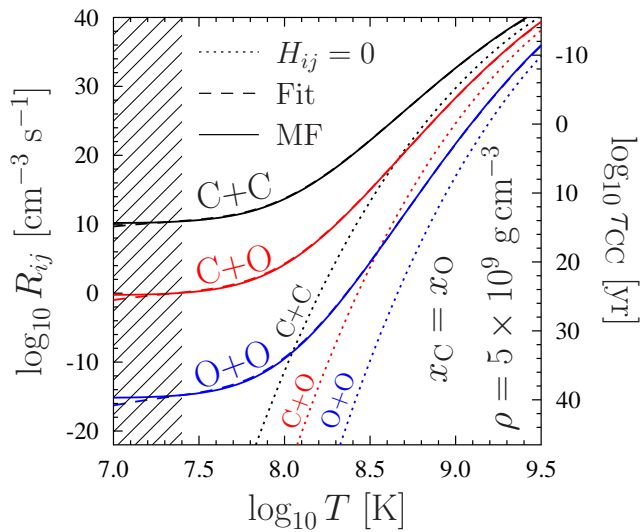}
    \end{center}
    \caption{(Color online) Reaction rates  in a
    $^{12}$C+$^{16}$O mixture with $x_\mathrm{C}=x_\mathrm{O}=0.5$
    at $\rho=5\times 10^9$~g~cm$^{-3}$.
    Solid lines are the mean-field calculations, dashed lines correspond to our fit
    expression (\ref{h_fit}), and dotted lines are obtained
    neglecting the plasma screening. The lines of the same
    type refer (from top to bottom) to
    the C+C, C+O, and O+O reactions.
    The shaded region of $T<2.5\times 10^{7}$~K
    corresponds to bound Gamow-peak states, $E^\mathrm{pk}_{ij}<0$.
    The right vertical scale shows characteristic carbon burning time
    $\tau_\mathrm{CC}=n_\mathrm{C}/R_\mathrm{CC}$.
    See text for details.}
    \label{fig:ReactRate}
\end{figure}

Figure \ref{fig:ReactRate} shows the temperature dependence of the
three (C+C, C+O, and O+O) reaction rates in a $^{12}\mathrm{C}$ and
$^{16}\mathrm{O}$ mixture with equal number density of carbon and
oxygen ions at $\rho=5\times 10^9$~g~cm$^{-3}$. The solid lines are
our mean-field calculations (marked as MF), the dashed lines are
given by the fit expression (\ref{h_fit}) (marked as Fit), and the
dotted lines are calculated neglecting the plasma screening
($H_{ij}(r)=0$). Lines of the same type refer (from top to bottom)
to the C+C, C+O, and O+O reactions, respectively. The right vertical
scale gives typical carbon burning time
$\tau_\mathrm{CC}=n_\mathrm{C}/R_\mathrm{CC}$. The shaded region
($T<2.5\times 10^{7}$~K) corresponds to bound Gamow-peak states (see
Fig.\ \ref{fig:gamow} and corresponding discussion in the text).
This region is similar for all three reactions because of
approximately the same charges of reacting ions. For such low
temperatures the mean field model with isotropic potential is
quantitatively invalid but qualitatively correct; the reaction rates
become temperature independent, which is the main property of
pycnonuclear burning. The pycnonuclear burning rates are rather
uncertain \cite{YGABW06}; typical rates reported in the literature
are one order of magnitude smaller than those extracted from our
mean field calculations at $T \to 0$. The difference may result from
the fact that we use spherically symmetric mean-field potential
(rather than more realistic anisotropic potential). In reality, ions
can be localized in deeper potential wells near their lattice sites.

The fit and mean-field results, which are mostly indistinguishable
in the figure, can strongly differ in the shaded region, especially
for the O+O reaction. We do not expect that it is a significant
disadvantage of our fit expression (because the mean field approach
becomes invalid at such conditions), but we would like to mention
this feature. Note, that for the O+O reaction in Fig.\
\ref{fig:ReactRate} the temperature $T=2.5\times 10^7$~K corresponds
to $\Gamma_\mathrm{OO}\approx400$ and the total enhancement factor
is $\sim 10^{110}$.

\begin{figure}
    \begin{center}
        \epsfxsize=68.7mm
        \epsfbox{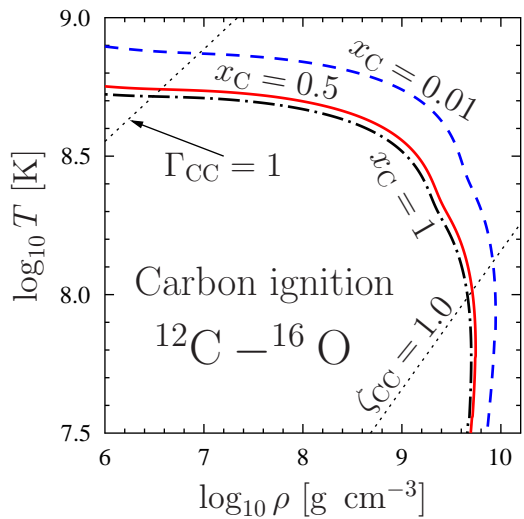}
    \end{center}
    \caption{(Color online) Carbon ignition curve in
    $^{12}$C+$^{16}$O mixtures.
    Dash-dotted, solid and dashed lines show the mean-field
    calculations for $x_\mathrm{C}=1$, $0.5$, and $0.01$,
    respectively.
    We also plot
    two thin dotted lines of constant $\Gamma_\mathrm{CC}=1$ and
    $\zeta_{\mathrm{CC}}=1$ to demonstrate typical values of dimensionless parameters.
    See text for details.}
    \label{fig:Ignition}
\end{figure}

Figure \ref{fig:Ignition} presents carbon ignition curves, which are
most important for modeling nuclear explosions of massive white
dwarfs (supernova Ia events) and carbon explosions in accreting
neutron stars (superbursts). For white dwarfs it is determined as a
line in the $T-\rho$ plane, where the nuclear energy generation rate
equals the local neutrino energy losses (which cool the matter). For
higher $T$ and $\rho$ (above the curve), the nuclear energy
generation exceeds the neutrino losses and carbon ignites. The
curves are plotted for $^{12}$C+$^{16}$O mixtures. The main energy
is generated in the C+C reaction even for $x_\mathrm{C}=0.01$
because the C+O and O+O reactions are stronger suppressed by the
Coulomb barrier (see Fig.\ \ref{fig:ReactRate} to compare the
reaction rates). In our mean-field model, which employs the linear
mixing, the admixture of oxygen affects the C+C burning only by
reducing the number density of carbon nuclei [not through the
contact probability $g_\mathrm{CC}(0)$; see Eq.\ (\ref{R_g_0})], so
that the C+C reaction rate is $\propto x_\mathrm{C}^2$. The neutrino
energy losses are mainly produced by plasmon decay and
electron-nucleus bremsstrahlung processes. The neutrino emissivity
owing to plasmon decay is calculated using the results of Ref.\
\cite{Plasmon} (with the online table
http://www.ioffe.ru/astro/NSG/plasmon/table.dat). The neutrino
bremsstrahlung emissivity is calculated using the formalism of
Kaminker {\it et al}.\ \cite{kaminkeretal99}, which takes into
account electron band structure effects in crystalline matter. For
an CO mixture, this neutrino emissivity is determined using the
linear mixture rule.

Two thin dotted lines in Fig.\ \ref{fig:Ignition} correspond to
constant $\Gamma_\mathrm{CC}=1$ and $\zeta_{\mathrm{CC}}=1$.
Dash-dotted, solid and dashed lines are the mean-field calculations
for $x_\mathrm{C}=1$ (pure carbon matter), $0.5$, and $0.01$,
respectively. For a fixed $\rho$, the higher $x_\mathrm{C}$ the
higher the number density of carbon ions and the higher the reaction
rate. This intensifies carbon burning, and the carbon ignites at
lower temperature. For high densities, the carbon ignition curve
bends and the carbon can ignite at very low temperatures. This is
caused by weakening the temperature dependence of the reaction rate
(see Fig.\ \ref{fig:ReactRate}) and by a strong suppression of
neutrino emission with decreasing temperature.

\section{Conclusions}
\label{sec:conc}


We have analyzed the plasma screening enhancement of nuclear
reaction rates in binary ionic mixtures. We have used a simple model
for the enhancement factor based on the radial WKB tunneling of the
reacting nuclei in their Coulomb potential superimposed with the
static mean-field potential created by neighboring plasma ions. We
have done accurate Monte Carlo calculations of the mean-field plasma
potential for a two-component strongly coupled plasma of ions and
proposed a simple and accurate analytic fit to the plasma potential
(Sec.\ \ref{meanfield}). We have calculated the plasma enhancement
factors of nuclear reaction rates in the mean-field WKB
approximation and have obtained their accurate fit (Sec.\
\ref{enh}). We have analyzed the effect of the plasma screening on
astrophysical $S$-factors and Gamow-peak energies (Sec.\
\ref{sec:Gamow}). To illustrate the results, we analyzed nuclear
burning in $^{12}$C+$^{16}$O mixtures (Sec.\ \ref{sec:res}).

We demonstrate that the mean-field WKB method gives qualitatively
correct (temperature independent) reaction rates even in the
zero-temperature pycnonuclear burning regime. In this regime the
dynamics of the reacting ions is determined by zero-point
vibrations; they fuse along selected (anisotropic) close-approach
trajectories \cite{svh69}; the mean-field radial WKB method was
initially expected to be absolutely inadequate. Let us mention in
passing that the problem of pycnonuclear burning has not been
accurately solved even for OCP plasma (e.g., see \cite{OCP_react}),
and the uncertainties of the solution increase in multicomponent
mixtures (see \cite{YGABW06}, and references therein).

Other uncertainties in our knowledge of the reaction rates come from
nuclear physics. As a rule, the astrophysical $S$-factors cannot be
experimentally measured for such low energies as Gamow peak energies
in stellar matter. For example, the lowest experimental point for
the $^{12}$C+$^{12}$C reaction is $\sim 2.1$~MeV, whereas typical
Gamow peak energies ${E^\mathrm{pk}_{ij}}^\prime$ are $\sim 1$~MeV.
Therefore, one needs to extrapolate experimental results to  lowest
energies. Throughout the paper we assumed a smooth energy dependence
of astrophysical $S$-factors, that is supported by calculations in
the frame of the barrier penetration model (e.g., Ref.\
\cite{Gasetal07}). However, some models (e.g., \cite{Perez_etal06}),
predict resonances at low energy, which can significantly change the
reaction rates and ignition curve \cite{Cooper_etal09}. We do not
discuss such effects in the present paper. New experimental and
theoretical studies of astrophysical $S$-factors are needed to solve
this problem.

\begin{acknowledgments}
We are grateful to D.G.~Yakovlev for useful remarks. Work of AIC was
partly supported by the Russian Foundation for Basic Research (grant
08-02-00837), and by the State Program ``Leading Scientific Schools
of Russian Federation'' (grant NSh 2600.2008.2). Work of HED was
performed under the auspices of the US Department of Energy by the
Lawrence Livermore National Laboratory under contract number
W-7405-ENG-48.
\end{acknowledgments}

\appendix


\section{Correction of the enhancement factors for weak nonideality}
\label{App_LowGamma}


\begin{figure*}
    \begin{center}
        \leavevmode
        \epsfxsize=0.95\textwidth \epsfbox{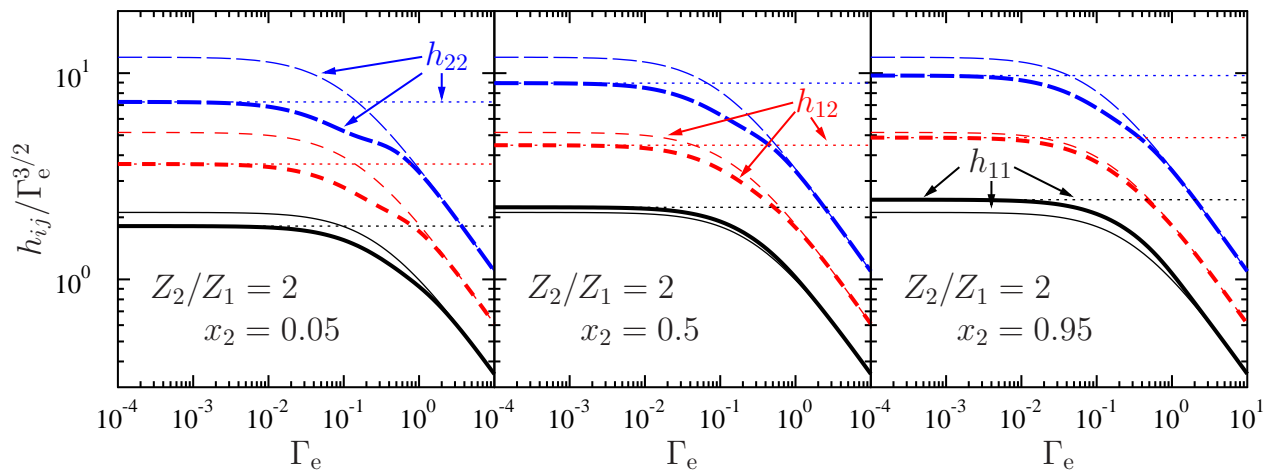}
    \end{center}
    \caption{(Color online) The enhancement coefficients
    $h_{ij}/\Gamma_\mathrm{e}^{3/2}$ (at small $\zeta_{ij}$)
    as a function of $\Gamma_\mathrm{e}$ for binary mixtures with
    $Z_2/Z_1=2$ and $x_2=0.05$ (left), $0.5$ (center) and $0.95$ (right).
    Thin dotted
    horizontal lines correspond to Debye-H\"{u}ckel enhancement
    (\ref{Debye_enh}). Thin solid, dash and long-dash lines are
    $h_{11}^\mathrm{lin}$, $h_{12}^\mathrm{lin}$, and
    $h_{22}^\mathrm{lin}$, respectively.
    Thick lines demonstrate the enhancement factors,
    calculated using our interpolation expression
    (\ref{h_fit_fin}).
    }
    \label{fig:LowGamma}
\end{figure*}

Our fit expression (\ref{h_fit}) does not reproduce the well-known
Debye-H\"{u}ckel enhancement factor for weak Coulomb coupling
($\Gamma_{ij}\ll 1$),
\begin{equation}\label{Debye_enh}
  h^\mathrm{DH}_{ij}=3^{1/2} Z_i\,Z_j\ZZmid^{1/2}\Gamma_e^{3/2}/\Zmid^{1/2}.
\end{equation}
Let us remind that $h_{ij}$ is related to the reaction rates by Eq.\
(\ref{h_def}). To correct our approximation in the weak-coupling
limit we note, that a formal use of the linear mixing rule [employed
in Eq.\ (\ref{h_fit})] at low $\Gamma_{ij}\ll1$ gives
\begin{equation}
  h^\mathrm{lin}_{ij}=3^{-1/2}\,\left[\left(Z_i+Z_j\right)^{5/2}-Z_i^{5/2}-Z_j^{5/2}\right]
  \Gamma_\mathrm{e}^{3/2}.
\end{equation}
Therefore, our fit (\ref{h_fit}) gives the correct power low
($\propto \Gamma_\mathrm{e}^{3/2}$), but an inexact prefactor (for
$0.1\le Z_1/Z_2\le 10$ and all $x_1$ the relative error does not
exceed 40\%). We suggest to introduce a correction factor
\begin{equation}
\label{LowGammaCorrection}
   C_{ij}=\frac{h^\mathrm{DH}_{ij}}{h^\mathrm{lin}_{ij}}
    =3Z_i\,Z_j
                 \frac{\ZZmid^{1/2}}
                         {\Zmid^{1/2}}
   \,\frac{1}{\left(Z_i+Z_j\right)^{5/2}-Z_i^{5/2}-Z_j^{5/2}},
\end{equation}
and finally write the enhancement factor as
\begin{equation}
\label{h_fit_fin}
    h_{ij}=\frac{C_{ij}+\Gamma_{ij}^2}
           {1+\Gamma_{ij}^2}
    \left[f_0\left(\frac{\Gamma_i}{\tau_{ij}}\right)
            +f_0\left(\frac{\Gamma_j}{\tau_{ij}}\right)
            -f_0\left(\frac{\Gamma_{ij}^\mathrm{comp}}{\tau_{ij}}\right)\right],
\end{equation}
where $\tau_{ij}$ is given by (\ref{h_fit_addpar}). As a result,
Eq.\ (\ref{h_fit_fin}) reproduces the correct Debye-H\"{u}ckel
asymptote of the enhancement factor in the weak coupling limit
($\Gamma_{ij}\ll 1$) and it reproduces also our results at
$\Gamma\gtrsim 1$.

Our interpolation expression is certainly a simplification, but it
is expected to be qualitatively correct. In Fig.\ \ref{fig:LowGamma}
we show the enhancement factors $h_{ij}$, normalized with respect to
$\Gamma_\mathrm{e}^{3/2}$, as a function of $\Gamma_\mathrm{e}$. We
take a BIM with $Z_2/Z_2$ and $x_2=0.05$, $0.5$ and $0.95$ as an
example. The thin dotted lines correspond to the Debye-H\"{u}ckel
asymptote at $\Gamma\ll1$. The thin solid, dash and long-dash lines
are the linear mixing asymptotes ($h_{11}^\mathrm{lin}$,
$h_{12}^\mathrm{lin}$, and $h_{22}^\mathrm{lin}$, respectively),
which are valid in the limit of strong Coulomb coupling,
$\Gamma\gg1$. The thick lines show the enhancement factors,
calculated using our interpolation expression (\ref{h_fit_fin}). One
can see that the asymptotes fix the enhancement factors quite well
and our interpolation looks reasonable.

Of course, an accurate description of the enhancement factors at
moderate Coulomb coupling is desirable but the exact solution may be
complicated. Fortunately, in all cases of physical interest the
reactions in this regime occur at $\zeta_{ij}\ll1$, so that we can
calculate the enhancement factors as a difference of free energies
before and after a reaction event. Recent calculations of Potekhin
et al.\ \cite{Potekhin_Mixt} of the free energy of binary and triple
mixtures at intermediate Coulomb coupling seem to be most important
in this respect. However, they are not very convenient for practical
purpose because the enhancement factors depend on derivatives of the
free energy with respect to numbers of particles. These derivatives
should be calculated at small concentrations of compound nuclei, and
the accuracy of numerical differentiation deserves a special study.

                    
\end{document}